\documentclass[a4paper,fleqn]{cas-dc}

\usepackage{amsmath, amssymb}
\usepackage[ruled,vlined]{algorithm2e}

\usepackage[authoryear]{natbib}
\usepackage{subfigure} 

\def\tsc#1{\csdef{#1}{\textsc{\lowercase{#1}}\xspace}}
\tsc{WGM}
\tsc{QE}
\tsc{EP}
\tsc{PMS}
\tsc{BEC}
\tsc{DE}

\hypersetup{hidelinks}
\begin{document}

\let\WriteBookmarks\relax
\def\floatpagepagefraction{1}
\def\textpagefraction{.001}
\shortauthors{Andrea Dosi et~al.}

\title [mode = title]{Less is More: AMBER-AFNO - a New Benchmark for Lightweight 3D Medical Image Segmentation}

\author[1]{Andrea Dosi}[orcid=0000-0002-5943-6867]
\cormark[1]
\ead{andrea.dosi@unina.it}

\author[1]{Semanto Mondal}[orcid=0009-0000-2306-4478]
\ead{semanto.mondal@unina.it}

\author[2]{Rajib Chandra Ghosh}[orcid=0009-0001-1137-3465]
\ead{rajib.chandraghosh@unina.it}

\author[1,3]{Massimo Brescia}[orcid=0000-0001-9506-5680]
\ead{massimo.brescia@unina.it}

\author[1]{Giuseppe Longo}[orcid=0000-0002-9182-8414]
\ead{giulongo@unina.it}

\affiliation[1]{organization={Department of Physics E. Pancini, University of Naples Federico II},
                addressline={Via Cinthia 21},
                city={Naples},
                postcode={80126},
                country={Italy}}

\affiliation[2]{organization={Department of Electrical Engineering and Information Technology (DIETI), University of Naples Federico II},
                addressline={Via Claudio 21},
                city={Naples},
                postcode={80125},
                country={Italy}}

\affiliation[3]{organization={INAF, Astronomical Observatory of Capodimonte},
                addressline={Salita Moiariello 16},
                city={Naples},
                postcode={80131},
                country={Italy}}

\cortext[cor1]{Corresponding author}

\begin{abstract}
We adapt the remote sensing-inspired AMBER model \citep{reference18} from multi-band image segmentation to 3D medical datacube segmentation. To address the computational bottleneck of the volumetric transformer, we propose the AMBER-AFNO architecture. This approach uses Adaptive Fourier Neural Operators (AFNO) instead of the multi-head self-attention mechanism. Unlike spatial pairwise interactions between tokens, global token mixing in the frequency domain avoids $\mathcal{O}(N^2)$ attention-weight calculations. As a result, AMBER-AFNO achieves quasi-linear computational complexity and linear memory scaling.

\noindent This new way to model global context reduces reliance on dense transformers while preserving global contextual modeling capability. By using attention-free spectral operations, our design offers a compact parameterization and maintains a competitive computational complexity. We evaluate AMBER-AFNO on three public datasets: ACDC, Synapse, and BraTS. On these datasets, the model achieves state-of-the-art or near-state-of-the-art results for DSC and HD95. Compared with recent compact CNN and Transformer architectures, our approach yields higher Dice scores while maintaining a compact model size.

\noindent Overall, our results show that frequency-domain token mixing with AFNO provides a fast and efficient alternative to self-attention mechanisms for 3D medical image segmentation.
\end{abstract}


\maketitle
\section{Introduction}\label{sec1}
\noindent According to the World Health Statistics 2024 report \citep{GlobalHealthEstimates}, heart disease was the leading cause of death globally in 2021, accounting for approximately 9.1 million deaths. Kidney disease, lung cancer, lower respiratory infections, and stroke also ranked among the top ten global causes of mortality. In all cases, early diagnosis is crucial to reduce mortality rates and optimize treatment strategies, including therapy or surgery. Such diagnosis relies on advanced imaging technologies such as MRI and CT scans, which generate three-dimensional (3D) volumetric data capturing anatomical structures at multiple depths \citep{reference3, florkow2022mri}. Since a single 2D slice often fails to capture the full anatomical context, effective analysis of volumetric data cubes has become essential \citep{ModelsGenesis}.

\noindent In medical image segmentation tasks, the U-Net architecture \citep{UNET} has gained widespread popularity due to its encoder–decoder structure, which captures both local and global contextual representations through skip connections. Numerous variants have been proposed to improve multi-scale representation and robustness, including AFFU-Net \citep{zheng2022affunet}, OAU-Net \citep{song2023oau}, ISTD-Net \citep{hou2022istdu}, MultiResU-Net \citep{lan2022dnn}, SAU-Net \citep{chen2024sau}, KiU-Net \citep{valanarasu2022kiunet}, and UCR-Net \citep{sun2022ucrnet}. However, convolutional neural networks (CNNs) inherently struggle to model long-range dependencies due to limited receptive fields determined by kernel size and stride, particularly in volumetric data where inter-slice relationships are critical.

\noindent To address these limitations, the Transformer \cite{vaswani2017attention} was introduced into computer vision as Vision Transformers (ViT) \cite{dosovitskiy2021image}. Transformer-based models rely on self-attention mechanisms to model long-range dependencies across the entire image. Consequently, various transformer-based architectures have been proposed for medical image segmentation, such as RT-UNet \cite{li2022rtunet}, SWTRU \cite{zhang2022star}, LMIS \cite{zhu2024lightweight}, and SDV-TUNet \cite{zhu2024sparse}, which combine transformer blocks with U-Net-like structures to capture both global and local information.

\noindent However, in volumetric settings, transformer-based models introduce a significant computational bottleneck. In 3D medical image segmentation, feature maps grow cubically with spatial resolution, and the quadratic complexity of token-to-token self-attention results in high memory consumption, increased parameter counts, and longer inference time. Although efficient attention mechanisms have been proposed to approximate self-attention, most still rely on token-wise interaction operations whose complexity scales poorly with high-resolution volumes. Therefore, designing a 3D segmentation model that balances effective global context modeling with computational efficiency remains an open problem.

\noindent SegFormer \citep{reference19} addresses efficiency in semantic segmentation through a lightweight hierarchical transformer encoder and an MLP-based decoder. Building upon this foundation, AMBER \citep{reference18} extends SegFormer to multi-band image segmentation by incorporating three-dimensional convolutions and a Funnelizer layer to better model spatial–spectral relationships.

\noindent To explicitly address the efficiency–globality trade-off in 3D lightweight medical image segmentation, we propose AMBER-AFNO, where Adaptive Fourier Neural Operators (AFNO) \citep{AFNO} replace multi-head self-attention within a hierarchical transformer encoder. Instead of constructing quadratic attention matrices, AFNO performs global token mixing in the frequency domain through learnable spectral filters, enabling quasi-linear computational complexity and linear memory cost. The multi-head structural design is preserved through block-wise frequency mixing, allowing different frequency subspaces to capture distinct semantic patterns without pairwise token interactions. In contrast to attention approximation strategies or convolution–attention compression schemes, this approach reformulates global context modeling itself rather than merely reducing attention computation.

\noindent Building on the AMBER architecture, the proposed AMBER-AFNO adapts spectral-domain token mixing to full 3D volumetric segmentation. This design reduces the parameter count by nearly 78\% compared with heavy transformer-based models such as UNETR++, while maintaining competitive FLOPs and segmentation performance. Unlike recent lightweight CNN–Transformer variants (e.g., LW-CTrans \citep{LW-CTrans}), whose efficiency is achieved by simplifying convolutional or attention modules, the lightweight property of AMBER-AFNO stems from spectral-domain global mixing.

\noindent We validate AMBER-AFNO on three public 3D medical segmentation benchmarks: ACDC \citep{ACDC}, Synapse \citep{Synapse}, and BraTS \citep{BRATS}. Using Dice Similarity Coefficient (DSC) and Hausdorff Distance (HD95) as evaluation metrics, the experimental results demonstrate that spectral-domain token mixing achieves a favorable accuracy–efficiency trade-off across diverse anatomical scenarios.

\section{Related Work}
\noindent Three-dimensional medical image segmentation has seen great progress since the adoption of encoder-decoder networks, such as U-Net and V-Net. Most recent studies can be roughly grouped into four non-mutually exclusive categories: (1) single-branch encoder-decoder networks, (2) hybrid and multi-branch encoders, (3) transformers and efficient attention models, and (4) lightweight 3D segmentation networks.

\noindent\textbf{Single-branch encoder–decoder networks.}

\noindent Many classical CNN-based segmentation networks have been proposed, many of which evolved from the original U-Net \citep{UNET}. Examples include UNet++ \citep{UNETR++}, Attention U-Net \citep{UNET}, CE-Net \citep{CE-Net}, DeepLab \citep{DeepLab}, and for volumetric data, V-Net \citep{v-net} that uses residual connections in 3D convolutional blocks and Dice loss, a loss function adapted to volumetric medical images. Later, nnU-Net \citep{reference14} demonstrated that adapting depth, patch size, and training settings to the dataset on the same architecture can lead to state-of-the-art performance on a multitude of tasks. Despite the high power of fully convolutional networks, they often use increasingly large receptive fields in deeper layers and thus cannot directly model long-range interactions across higher-resolution volumes.

\noindent\textbf{Hybrid and multi-branch encoders.}

\noindent CNN–Transformer frameworks were proposed to address locality limitations. TransUNet \citep{TransUNet} incorporated ViT modules into the CNN encoder, and UNETR \citep{UNETR} redefined U-Net with a transformer encoder and convolutional decoder. CoTr \citep{CoTr} introduced deformable attention in 3D space to reduce computational burden. Subsequent works designed multi-branch encoders to disentangle semantic resolutions or image details. DS-TransUNet \citep{DS-TransUNet} employed dual-scale ViT encoders; DHR-Net \citep{DHRNet} introduced semantic and detail branches; MILU-Net \citep{MILU-Net} and BSC-Net \citep{BSC-Net} adopted dual-resolution designs. However, multi-branch models usually bring significantly more parameters and increase training complexity, limiting scalability.

\noindent\textbf{Transformer-based and efficient attention models.}

\noindent Global self-attention in Vision Transformers has proven effective for representation learning. In volumetric segmentation, UNETR \citep{UNETR} and Swin-UNETR \citep{swin_unetr} demonstrate the importance of hierarchical token interactions. nnFormer \citep{nnFormer} alternates convolution and attention blocks, and UNETR++ \citep{UNETR++} proposed Efficient Paired Attention (EPA) for separating spatial and channel attention. DS-UNETR++ \citep{DS-UNETR++} further introduced gated cross-attention mechanisms.

\noindent In its standard form, self-attention grows quadratically with the number of tokens, making it costly for larger 3D inputs. Linformer \citep{Linformer} and Performer \citep{Performers} apply low-rank and kernel-based approximations, while FNet \citep{FNet} replaces attention with fixed Fourier transforms.

\noindent In contrast, Adaptive Fourier Neural Operators (AFNO) \citep{AFNO} learn frequency filters and block diagonal channel mixing matrices for token mixing in the Fourier domain. By truncating high-frequency modes and applying adaptive frequency filters, AFNO achieves quasi-linear complexity and linear memory cost while modeling global context. AFNO was originally proposed for 3-channel images, and its potential for volumetric medical segmentation remains underexplored.

\noindent\textbf{Lightweight 3D medical image segmentation networks.}

\noindent For real-world clinical applications where computational resources are limited, efficient 3D segmentation aims to reduce parameters and floating-point operations while maintaining competitive performance. Efficient design principles from 2D image processing, such as depthwise separable convolution and inverted residual blocks (MobileNetV2 \citep{MobileNetV2}), and grouped pointwise convolution with channel shuffle (ShuffleNet \citep{ShuffleNet}), have inspired lightweight 3D models.

\noindent LCOVNet \citep{LCOV-NET} utilizes separable spatiotemporal convolution with attention calibration, and ADHDC-Net \citep{ADHDC-Net} employs decoupled convolution to replace standard convolution in medical images. SlimUNETR \citep{Slim_UNETR} simplifies the attention mechanism, and LW-CTrans \citep{LW-CTrans} combines lightweight convolution encoding with simplified transformer modules to significantly reduce parameter count while maintaining comparable Dice scores.

\noindent Most lightweight variants focus on convolutional filter redesign or attention simplification, with limited investigation of token mixing strategies themselves. In particular, spectral and domain-agnostic variants such as AFNO have not been systematically applied to volumetric medical image segmentation. Consequently, whether accurate 3D segmentation can be achieved with significantly reduced parameter counts compared to current state-of-the-art methods remains unclear.

\smallskip

\noindent In summary, volumetric medical image segmentation architectures have evolved from CNN encoder–decoders to CNN–Transformer hybrids and fully attention-based models emphasizing global context modeling. While transformer-based approaches enhance long-range dependencies, their quadratic attention mechanisms are computationally intensive for high-resolution 3D images.

\noindent Unlike the above approaches, \textit{AMBER-AFNO} bypasses pairwise attention mechanisms by performing global mixing in the frequency domain. By applying Fourier transforms and learning frequency filters over limited modes, AMBER-AFNO avoids constructing $\mathcal{O}(N^2)$ attention matrices and enables quasi-linear complexity with linear memory usage, while significantly reducing parameter counts and preserving the ability to handle high-resolution 3D volumes.

\section{Methodology}\label{sec3}
\begin{figure*}
	\centering
	\includegraphics[width=\textwidth]{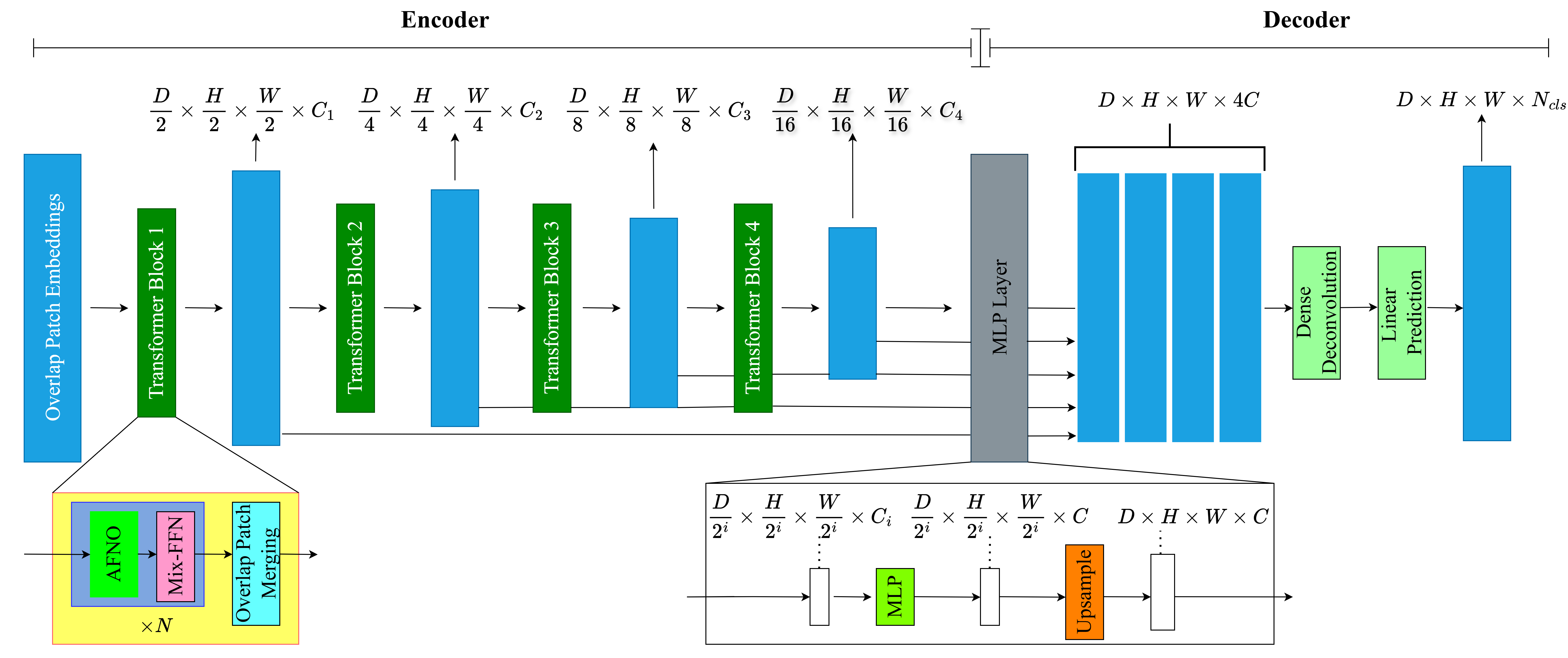}
	\caption{\textbf{The Proposed AMBER-AFNO framework} consists of two main modules: A hierarchical Transformer encoder to extract coarse and fine features; and a lightweight MLP decoder to directly fuse these multi-level features and predict the semantic segmentation mask. \textbf{FFN} indicates a feed-forward network.}
	\label{The Proposed AMBER-AFNO framework}
\end{figure*}

\noindent In this section, we introduce AMBER-AFNO, an extension of the AMBER model \citep{reference18}, which replaces the traditional self-attention mechanism with Adaptive Fourier Neural Operators (AFNO) to reduce model complexity and improve computational efficiency. While staying close to the original AMBER design, AMBER-AFNO adapts the architecture to address the 3D volumetric data. As illustrated in Figure~\ref{The Proposed AMBER-AFNO framework}, the architecture consists of two main modules: a hierarchical Transformer encoder with 3D patch embedding and AFNO-based feature mixing; and a lightweight MLP decoder that fuses multi-scale features and predicts the final 3D segmentation mask. Unlike AMBER, which uses a dimensionality reduction layer (“\textit{funnalizer}”) to collapse 3D features into 2D outputs, AMBER-AFNO operates entirely in 3D and directly outputs a volumetric segmentation mask of size $D \times H \times W \times N_{cls}$, where $N_{cls}$ is the number of classes. Furthermore, the decoder integrates a deconvolutional layer to up-sample feature maps and recover the original spatial resolution.

\subsection{Hierarchical Transformer Encoder}
\label{Hierarchical Transformer Encoder}
\noindent We designed a series of Mix Transformer encoders (MiT) for semantic segmentation of 3D images, replacing standard self-attention layers with Adaptive Fourier Neural Operators (AFNO) to reduce the number of parameters without compromising accuracy. Unlike conventional attention mechanisms—which are memory-intensive and scale quadratically with input size—AFNO performs global feature mixing in the frequency domain with quasi-linear complexity. This enables the encoder to capture both local and global context while significantly lowering computational and memory costs.
\newline
\textbf{Hierarchical Feature Representation.} 
The goal of this module is, given an input image, to generate CNN-like multi-level features. These features provide high-resolution coarse features and low-resolution fine-grained features that usually boost the performance of semantic segmentation \citep{reference19}. More precisely, given an input 3D image with $D \times H \times W$, we perform patch merging to obtain a hierarchical feature $F_{1}$ with a resolution of $\frac{D_{qw}}{2^i} \times \frac{H}{2^i} \times \frac{W}{2^i} \times C_1$, where $i \in \{0, 1, 2, 3\}$ and $C_{i+1}$ is larger than $C_{i}$.

\noindent \textbf{Overlapped Patch Merging.} 
 We utilize merging overlapping patches to avoid the need for positional encoding. To this end, we define $K$, $S$, and $P$, where $K$ is the three-dimensional kernel size (or patch size), $S$ is the stride between two adjacent patches, and $P$ is the padding size. Unlike the original SegFormer, in our experiments, we set $K$ = 3, $S$ = 1, $P$ = 1, and $K$ = 3, $S$ = 2, $P$ = 2 to perform overlapping patch merging. The patch size is intentionally kept small to preserve image details and avoid parameter explosion. $S$ = 1 preserves the original image spatial dimensions $H$ and $W$, avoiding the reduction of spatial dimensions by 1/4.

\noindent \textbf{Adaptive Fourier Neural Operators (AFNO).} 
The self-attention mechanism, while effective in capturing global dependencies, is widely recognized as the primary computational bottleneck in transformer-based encoder architectures due to its quadratic complexity with respect to the input sequence length \citep{reference19}. \newline
In AMBER \citep{reference18}, a multi-head self-attention scheme is used, where each attention layer incorporates a reduction factor $R$ to mitigate this cost. This reduces the computational complexity from the standard $O(N^2)$ to $O\left(\frac{N^2}{R}\right)$, offering a more tractable solution for high-resolution inputs \citep{reference19}.

\noindent In the proposed AMBER-AFNO architecture, the entire attention mechanism is replaced with the AFNO block. AFNO uses token mixing in the frequency domain, leveraging the Fast Fourier Transform (FFT) to achieve global interactions with quasi-linear complexity. Here, the input tokens are transformed into the frequency domain using FFT, capturing global contextual information efficiently. \newline 
In \citep{AFNO} the authors introduced AFNO for the 2D image segmentation task. Using similar methodologies, we have extended the approach for the 3D image segmentation task. 

\noindent The input to the AFNO block is a 5-D tensor which can be represented as $
x \in \mathbb{R}^{B \times D \times H \times W \times C}$
where $B$ is the batch size, $D$, $H$, and $W$ are the spatial dimensions (depth, height, and width), and $C$ is the embedding dimension.

\noindent We then apply a real-valued 3D Fast Fourier Transform (RFFT) over the spatial dimensions:
\begin{equation}
\hat{x} = \text{RFFT}_3(x) \in \mathbb{C}^{B \times D \times H \times (W/2+1) \times C}
\end{equation}

\noindent The channel dimension $C$ is partitioned into $K$ frequency blocks:
\begin{equation}
\hat{x} \rightarrow \hat{x}_{\text{blk}} \in \mathbb{C}^{B \times D \times H \times (W/2+1) \times K \times \frac{C}{K}}
\end{equation}

\noindent Each frequency block undergoes a learnable complex-valued two-layer MLP:
\begin{equation}
\hat{x}_{\text{blk}}^{(i)} \leftarrow W_2^{(i)} \cdot \phi\left(W_1^{(i)} \cdot \hat{x}_{\text{blk}}^{(i)} + b_1^{(i)}\right) + b_2^{(i)}, \quad \forall i \in \{1, \dots, K\}
\end{equation}

\noindent After the operation of two layer MLP, the MLP output is reshaped back to $\hat{x} \in \mathbb{C}^{B \times D \times H \times (W/2+1) \times C}$ and then soft shrinkage is applied to attenuate small-magnitude frequency responses:
\begin{equation}
\hat{x}' = \text{SoftShrink}(\hat{x})
\end{equation}

\noindent Finally, the inverse 3D FFT (IRFFT) is applied, and the result is combined with the original input via residual addition:
\begin{equation}
\tilde{x} = \text{IRFFT}_3(\hat{x}') + x
\end{equation}

\noindent Thus, the final output of the AFNO-3D block is as follows:
\begin{equation}
\boxed{
\tilde{x} = \text{IRFFT}_3\left( \text{SoftShrink}\left( \text{MLP}_{\mathbb{C}} \left( \text{RFFT}_3(x) \right) \right) \right) + x
}
\end{equation}

\begin{algorithm}[htbp]
\DontPrintSemicolon
\caption{AFNO-3D with Adaptive Weight Sharing and Adaptive Masking}
\label{alg:afno3d}

\KwIn{Tensor $x \in \mathbb{R}^{b \times d \times h \times w \times c}$}
\KwOut{Transformed tensor $y$}

\BlankLine
$bias \leftarrow x$\;
$x \leftarrow \text{RFFT3}(x)$\;
$x \leftarrow \text{reshape}(x, [b, d, h, \frac{w}{2}+1, k, \frac{c}{k}])$\;
$x \leftarrow \text{BlockMLP}(x)$\;
$x \leftarrow \text{reshape}(x, [b, d, h, \frac{w}{2}+1, c])$\;
$x \leftarrow \text{SoftShrink}(x)$\;
$x \leftarrow \text{IRFFT3}(x)$\;
\Return $x + bias$\;

\BlankLine
\SetKwFunction{BlockMLP}{BlockMLP}
\SetKwProg{Fn}{Function}{:}{}
\Fn{\BlockMLP{$x$}}{
    $x \leftarrow x W_1 + b_1$\;
    $x \leftarrow \text{ReLU}(x)$\;
    \Return $x W_2 + b_2$\;
}
\end{algorithm}



\noindent \textbf{Algorithm} \ref{alg:afno3d} shows the pseudo-implementation of the AFNO-3D block in the AMBER-AFNO architecture. \newline
\textbf{Mix-FFN.} 
\vspace{0.5mm}
Likewise, in the AMBER \citep{reference18} and SegFormer \citep{reference19}, we also used the Mix-FFN, which considers the effect of zero padding using a $3 \times 3 \times 3$ Conv in the feed-forward network (FFN). We used Mix-FFN instead of positional encoding because Mix-FFN combines both Depthwise Convolution and MLP layers to capture both local and global context in the seen \citep{reference19}.
\begin{equation}
x_{\text{out}} = \text{MLP}\left(\text{GELU}\left(\text{Conv}_{3\times3\times3}\left(\text{MLP}(x_{\text{in}})\right)\right)\right) + x_{\text{in}}
\end{equation}
where $x_{in}$ is the feature from the self-attention module. Mix-FFN mixes a $3 \times 3 \times 3$ convolution and an MLP into each FFN.
\subsection{Lightweight All-MLP Decoder}
\label{Lightweight All-MLP Decoder}    
\noindent The four feature maps produced by the MiT encoder are first channel-projected to a common embedding size \(d\) by position-agnostic MLP layers.  Then every projected tensor is trilinearly upsampled to the finest encoder resolution and concatenated along the channel axis.  A \(1\times1\times1\) convolution with \textsc{ReLU} activation and 3-D batch normalization fuses this aggregate into a compact representation.  A single transposed 3-D convolution subsequently doubles the spatial dimensions, bringing the volume back to the native voxel grid while reducing the channels to \(N_{\mathrm{cls}}\).  A final \(1\times1\times1\) convolution sharpens the logits, yielding the segmentation mask  
\(M\in\mathbb{R}^{N_{\mathrm{cls}}\times D\times H\times W}\).  Compared with the original five-stage AMBER decoder, this variant omits the dedicated spectral-reduction block and integrates the final MLP into the transposed convolution, lowering memory cost without sacrificing accuracy.

\begin{figure}
	\centering
	\includegraphics[width=\columnwidth]{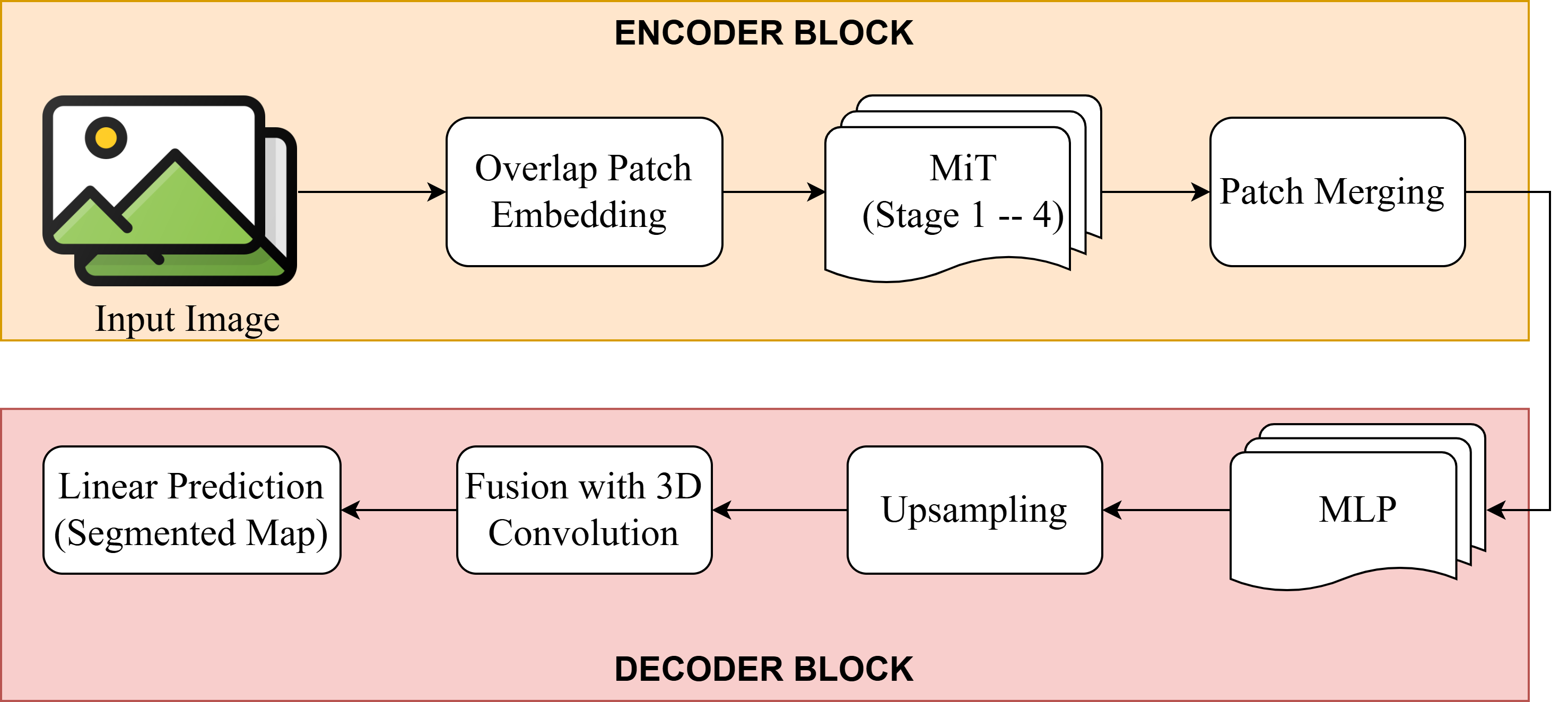}
	\caption{Simplified Work Flow Diagram of AMBER-AFNO Architecture}
        \label{Simplified Work Flow Diagram of AMBER 3D Architecture}
\end{figure}

\noindent The figure \ref{Simplified Work Flow Diagram of AMBER 3D Architecture} shows a more simplified and straightforward workflow diagram of the AMBER-AFNO Architecture.

\section{Experiments} \label{Data_Description}
\noindent To assess the effectiveness of the proposed \textsc{AMBER-AFNO} architecture, we compare against the following state-of-the-art convolutional and transformer-based segmentation models: U-Net \citep{UNET}, nnU-Net \citep{reference14}, TransUNet \citep{TransUNet}, Swin-UNet \citep{Swin-Unet}, UNETR \citep{UNETR}, UNETR++ \citep{UNETR++}, MISSFormer \citep{MISSFormer}, Swin-UNETR \citep{swin_unetr}, nnFormer \citep{nnFormer}, TransBTS \citep{TransBTS}, CoTr \citep{CoTr}, PCCTrans \citep{PCCTrans}, and LW-CTrans \citep{LW-CTrans}. \newline
The last one mentioned, LW-CTrans \citep{LW-CTrans}, is a newly introduced lightweight 3D CNN-Transformer hybrid that shares the same goal as our approach (i.e., fewer model parameters) for volumetric segmentation. This baseline is presented to not only evaluate the performance of our method against the mainstream choice of a lightweight 3D CNN-Transformer hybrid but also test whether the proposed frequency-domain token mixing (AFNO) can deliver a better trade-off between performance and efficiency, given a similar compact model size constraint.\newline
Although it is not comprehensive, this list offers a competitive and diverse set of solid baselines. These cover purely convolutional, convolutional-recurrent, attention-based, and lightweight models.\newline
All models are tested on three publicly available datasets: the ACDC challenge dataset for cardiac MR image segmentation, the Synapse dataset for multi-organ abdominal CT segmentation, and the BraTS dataset. The next three sub-sections describe the datasets and data pre-processing steps, the training and inference protocols used in the study, and the two quantitative metrics (HD95 and DSC) used for comparative analysis.

\subsection{Dataset Overview}
\label{Detailed Description}
\subsection*{Automated cardiac diagnostic segmentation dataset (ACDC)}
\noindent The ACDC dataset \citep{ACDC} consists of 3D cardiac MRI images with multi-class annotations. A total of 200 labeled samples were used, split into 160 for training and 40 for testing. The annotated classes include the right ventricle (RV), myocardium (MYO), and left ventricle (LV). In this study, the Dice Similarity Coefficient (DSC) was used as the evaluation metric for model comparison.
\empty
\subsection*{Multi-organ CT Segmentation Dataset (Synapse)}
\noindent The Synapse dataset \citep{Synapse} contains 30 abdominal CT scans and is commonly used for benchmarking multi-organ segmentation methods. Following the UNETR++ preprocessing protocol \citep{UNETR++}, the dataset is split into 18 training volumes and 12 validation volumes. It provides annotations for eight abdominal organs: spleen, right kidney, left kidney, gallbladder, liver, stomach, aorta, and pancreas.
For quantitative evaluation, we adopt two standard metrics in medical image segmentation: the Dice Similarity Coefficient (DSC) to measure volumetric overlap, and the 95th percentile Hausdorff Distance (HD95) to assess boundary accuracy.
\empty
\subsection*{Brain Tumor Segmentation (BraTS)}
\noindent The BraTS dataset \citep{BRATS} contains multimodal 3D brain MRI scans of patients with gliomas, which is a type of brain tumor. It contains 387 MRI scan images for training and 73 scans for testing. Here, each image consists of four channels, such as FLAIR, T1w, T1gd, and T2w. The images have four classes, including whole tumor (WT), enhancing tumor (ET), tumor core (TC) and background. In this experiment, the Dice Similarity Coefficient (DSC) is used to measure volumetric overlap, and the 95th percentile Hausdorff Distance (HD95) to assess boundary accuracy.

\begin{table*}[t]
\caption{High-level description of the datasets.}
\label{tab:datasets}
\begin{tabular*}{\textwidth}{@{\extracolsep{\fill}} p{0.10\textwidth} p{0.15\textwidth} p{0.15\textwidth} p{0.12\textwidth} p{0.08\textwidth} p{0.15\textwidth} p{0.13\textwidth}}
\toprule
Dataset & Spatial Dimension & Depth Dimension & Modality & Class & Training Samples & Test Samples \\
\midrule
ACDC & 320 × 320 (resampled) & 90–130 slices & 1 (Cine MRI) & 4 & 160 cases & 40 cases \\
Synapse & 512 × 512 & 75–250 slices & 1 (CT) & 9 & 18 scans & 12 scans \\

Brain Tumor (BraTS) & 240 × 240 & 155 slices & 4 (MRI) & 4 & 387 scans & 73 scans \\

\bottomrule
\end{tabular*}
\end{table*}
\section{Experimental Settings}\label{Experimental settings}
\noindent
For a fair comparison with existing transformer-based baseline methods, we follow the same preprocessing and training scheme as UNETR++ \citep{UNETR++}, UNETR \citep{UNETR}, DS-UNETR++ \citep{DS-UNETR++} and SAM-based medical segmentation models \citep{sam3d}. For each dataset, we use the same preprocessing procedures, including resampling to the same voxel spacing, intensity normalization, Z-score normalization, and the same cropping and padding protocols. No additional training data are used.\newline
All models are trained using the same input resolutions, as reported in UNETR++, to guarantee strict comparability. For Synapse, the input size is $128 \times 128 \times 64$; for BTCV, $96 \times 96 \times 96$; for ACDC, $160 \times 160 \times 16$ and for BraTS, $128 \times 128 \times 128$. All other data augmentation strategies and optimization details follow nnFormer \citep{nnFormer} to maintain consistency across baselines.\newline
Training is conducted on a single NVIDIA Tesla V100-SXM2 32GB GPU (300W TDP) using CUDA 12.2. All models are trained for 1,000 epochs with an initial learning rate of $0.01$ and weight decay of $3 \times 10^{-5}$. During inference, we employ a sliding-window strategy with 50\% overlap and report the Dice score of a single model without ensemble techniques. \newline
Deep supervision~\citep{Li2022} is employed during training of \textit{AMBER-AFNO} on the ACDC and Synapse datasets, thereby improving convergence stability and segmentation accuracy by enforcing auxiliary supervision at intermediate decoder stages. In contrast, for the BraTS dataset, empirical validation showed that removing deep supervision improved performance; thus, the reported BraTS results are from a model trained without deep supervision. Our loss function is a sum of Dice and Cross-Entropy losses, the same loss function used for UNETR++ and other transformer-based baseline models. A detailed formulation of the loss function is provided in the following section.

\section{Loss Function}\label{Loss Function}
\noindent For the semantic segmentation task, one possible problem that arises is class imbalance. We have considered three datasets with multilabel segmentation problems. All the datasets have class imbalance issues. The background class dominates over other classes, as the ROI for medical images is very low compared to the original shape of the image. One loss function alone, such as focal loss, dice loss or cross entropy loss can not handle this situation. To address this issue we have used a custom weighted loss function using the concept of Deep Supervision, which is a combination of cross-entropy loss and dice loss. 
Instead of relying on the final output of the model, the intermediate feature maps or predictions at different resolutions of the decoder block are also considered while calculating the loss and backpropagation.

\begin{equation}
    \begin{aligned}
L(G, P) = 1 - & \frac{2}{J} \sum_{j=1}^{J} \frac{ \sum_{i=1}^{I} G_{i,j} P_{i,j} }
{ \sum_{i=1}^{I} G_{i,j}^2 + \sum_{i=1}^{I} P_{i,j}^2 } \\
& - \frac{1}{I} \sum_{i=1}^{I} \sum_{j=1}^{J} G_{i,j} \log P_{i,j}
    \end{aligned}
\end{equation}
Where $\mathbf{G}$ refers to the set of ground truth labels, and $\mathbf{P}$ refers to the set of predicted probabilities. $P_{i,j}$ and $G_{i,j}$ represent the predicted probability and the one-hot encoded true value of class $j$ at voxel $i$, respectively. $I$ denotes the total number of voxels, and $J$ denotes the number of classes \citep{DS-UNETR++}.

\section{Evaluation Metrics}
\label{Evaluation Metrics}
\noindent We have adopted two primary evaluation metrics to assess segmentation performance: the Dice Similarity Coefficient (DSC), which quantifies the overlap between predicted and ground truth regions, and the 95th percentile Hausdorff Distance (HD95), which measures the spatial distance between boundary surfaces while mitigating the influence of outliers. Detailed definitions and computation procedures for these metrics are provided in the following subsection.
\subsection*{Hausdorff Distance (HD95)}
\noindent The HD 95 is a boundary-based metric that evaluates segmentation quality by computing the 95th-percentile distance between the predicted volume’s boundary voxels and those of the ground-truth segmentation.
\begin{equation}
HD_{95}(Y,P) \;=\; \max\bigl(d_{95}(Y,P),\, d_{95}(P,Y)\bigr)
\end{equation}
Here, $d_{95}(Y,P)$ is the maximum 95th percentile distance
between the ground truth and predicted voxels, and $d_{95}(P, Y )$
is the maximum 95th percentile distance between the predicted
and ground truth voxels.
\subsection*{Dice Similarity Coefficient (DSC).}
\noindent The Dice Similarity Coefficient (DSC) measures the similarity between two sets, returning values from 0 to 1, with 1 representing perfect similarity. It is computed using the following formula:
\begin{equation}
DSC(G,P)
  = \frac{2\,|G \cap P|}{|G| + |P|}
  = \frac{2 \sum_{i=1}^{I} G_i P_i}{\sum_{i=1}^{I} G_i + \sum_{i=1}^{I} P_i}
\end{equation}
Where \( G \) is the set of real results, \( P \) refers to the set of predicted results, \(G_{i}\) and \(P_{i}\) represent the true and predicted values of the voxel \textit{i}, respectively, and \textit{I }is the number of voxels.

\section{Results}\label{Result}
\noindent In this section, we evaluate \textsc{AMBER-AFNO} on three benchmark datasets: ACDC, Synapse, and BraTS. Performance is compared against state-of-the-art CNN and transformer models, including U-Net \citep{UNET}, nnU-Net \citep{reference14}, TransUNet \citep{TransUNet}, Swin-UNet \citep{Swin-Unet}, UNETR \citep{UNETR}, UNETR++ \citep{UNETR++}, MISSFormer \citep{MISSFormer}, Swin-UNETR \citep{swin_unetr}, nnFormer \citep{nnFormer}, TransBTS \citep{TransBTS}, CoTr \citep{CoTr}, PCCTrans \citep{PCCTrans} and LW-CTrans \citep{LW-CTrans}.\newline
In particular, LW-CTrans \citep{LW-CTrans} is a lightweight CNN-Transformer baseline designed to reduce parameter count and computational cost while maintaining competitive segmentation accuracy. Its baseline comparison makes it convenient to validate whether our proposed token-mixing approach in the frequency domain can achieve the same or higher segmentation accuracy while maintaining a lightweight design. \newline
We will perform quantitative (Dice, HD95) and qualitative (visual) assessment of the accuracy of the segmentations and contours, as well as the anatomical variability between the datasets.
\subsection{ACDC Dataset}
\noindent As shown in Tab.~\ref{tab:acdc_results} and Tab.~\ref{tab:acdc_params}, \textsc{AMBER-AFNO} ranks first in the overall Dice score (92.85 ) on the ACDC validation set, with an improvement over UNETR++ (92.83 ), and significantly fewer parameters (14.77M vs. 66.8M). With nearly 4 times fewer parameters, the proposed model still outperforms UNETR++ in segmentation accuracy, validating that token mixing in the spectral domain can effectively eliminate redundancy in the network.\newline
In contrast to the latest CNN-Transformer model, LW-CTrans (92.62\%), AMBER-AFNO achieves a higher Dice score with lower computational complexity (163.27G vs. 275.49G FLOPs). It suggests that frequency-domain token mixing can yield more effective global context modeling than compressed attention-based hybrid models under similar lightweight settings.
At the class level, our approach yields the highest Dice score for the myocardium (90.74) and the second-highest for the right (91.60) and left (96.21) ventricles, demonstrating its overall balanced performance across all cardiac classes.\newline
These results show that, on the ACDC dataset, the AFNO-based encoder coupled with a lightweight SegFormer-style decoder delivers state-of-the-art Dice performance while maintaining a favorable accuracy–efficiency trade-off compared to both heavy transformer-based models and recent lightweight CNN–Transformer architectures.

\begin{table*}[ht]
\centering
\caption{Dice scores (\%) for the right ventricle (RV), myocardium (Myo), and left ventricle (LV), together with the mean DSC (\%) on the ACDC validation set. \textbf{Bold} values denote the best result in each column, while \underline{underlined} values denote the second best.}
\label{tab:acdc_results}
\begin{tabular*}{\textwidth}{@{\extracolsep{\fill}}lcccc}
\toprule
\textbf{Methods} & \textbf{RV} & \textbf{Myo} & \textbf{LV} & \textbf{DSC} \\
\midrule
TransUNet\citep{TransUNet}       & 88.86 & 84.54 & 95.73 & 89.71 \\
Swin-UNet\citep{Swin-Unet}       & 88.55 & 85.62 & 95.83 & 90.00 \\
UNETR\citep{UNETR}               & 85.29 & 86.52 & 94.02 & 86.61 \\
MISSFormer\citep{MISSFormer}     & 86.36 & 85.75 & 91.59 & 87.90 \\
nnFormer\citep{nnFormer}         & 90.94 & 89.58 & 95.65 & 92.06 \\
UNETR++\citep{UNETR++}           & \textbf{91.89} & \underline{90.61} & 96.00 & \underline{92.83} \\
PCCTrans\citep{PCCTrans}         & 90.55 & 90.57 & \underline{96.22} & 92.45 \\
LW-CTrans \citep{LW-CTrans}      & 91.07 & 90.50 & \textbf{96.24} & 92.62 \\
\midrule
\textbf{AMBER--AFNO (ours)}      & \underline{91.60} & \textbf{90.74} & {96.21} & \textbf{92.85} \\
\bottomrule
\end{tabular*}
\end{table*}

\begin{table*}[ht]
\centering
\caption{ Comparison on ACDC. AMBER-AFNO achieves best segmentation results(DSC), while being efficient (Params in millions)}
\label{tab:acdc_params}
\begin{tabular*}{\textwidth}{@{\extracolsep{\fill}}lcccc}
\toprule
\textbf{Methods} & \textbf{Params} & \textbf{ FLOPs} & \textbf{DSC} \\
\midrule
UNETR++\citep{UNETR++}           & 66.8 & 43.71 & 92.83 \\
LW-CTrans \citep{LW-CTrans}          & 4.42 & 275.49 & 92.62 \\
\midrule
\textbf{AMBER--AFNO (ours)}      & {14.77 } & {163.27} & \textbf{92.85} \\
\bottomrule
\end{tabular*}
\end{table*}

\begin{table*}[ht!]
\centering
\caption Dice scores (\%) for eight abdominal organs and HD95 (mm) on the Synapse validation set. \textbf{Bold} values denote the best result in each column, while \underline{underlined} values denote the second best.
\label{tab:synapse_results}
\resizebox{\textwidth}{!}{%
\begin{tabular}{lcccccccccc}
\toprule
\textbf{Methods} & \textbf{Spl} & \textbf{RKid} & \textbf{LKid} & \textbf{Gal} & \textbf{Liv} & \textbf{Sto} & \textbf{Aor} & \textbf{Pan} & \textbf{HD95} & \textbf{DSC} \\ 
\midrule
U\hbox{-}Net\citep{UNET}            & 86.67 & 68.60 & 77.77 & 69.72 & 93.43 & 75.58 & 89.07 & 53.98 & --    & 76.85 \\
TransUNet\citep{TransUNet}          & 85.08 & 77.02 & 81.87 & 63.16 & 94.08 & 75.62 & 87.23 & 55.86 & 31.69 & 77.49 \\
Swin\hbox{-}UNet\citep{Swin-Unet}   & 90.66 & 79.61 & 83.28 & 66.53 & 94.29 & 76.60 & 85.47 & 56.58 & 21.55 & 79.13 \\
UNETR\citep{UNETR}                  & 85.00 & 84.52 & 85.60 & 56.30 & 94.57 & 70.46 & 89.80 & 60.47 & 18.59 & 78.35 \\
MISSFormer\citep{MISSFormer}        & 91.92 & 82.00 & 85.21 & 68.65 & 94.41 & 80.81 & 86.99 & 65.67 & 18.20 & 81.96 \\
nnFormer\citep{nnFormer}            & 90.51 & 86.25 & 86.57 & \underline{70.17} & \textbf{96.84} & \textbf{86.83} & \underline{92.04} & \textbf{83.35} & 10.63 & \underline{86.57} \\
Swin\hbox{-}UNETR\citep{swin_unetr} & \underline{95.37} & \underline{86.26} & 86.99 & 66.54 & 95.72 & 77.01 & 91.12 & 68.80 & \underline{10.55} & 83.48 \\
UNETR++\citep{UNETR++}              & \textbf{95.77} & \textbf{87.18} & \textbf{87.54} & \textbf{71.25} & \underline{96.42} & \underline{86.01} & \textbf{92.52} & \underline{81.10} & \textbf{7.53}  & \textbf{87.22} \\
PCCTrans\citep{PCCTrans}            & 88.84 & 82.64 & 85.49 & 68.79 & 93.45 & 71.88 & 86.59 & 66.31 & 17.10 & 80.50 \\
LW-CTrans \citep{LW-CTrans}         & 89.48 & 82.84 & 85.11 & 37.52 & 93.60 & 63.10  & 81.75 & 53.30 & 31.47 & 73.34 \\
\midrule
\textbf{AMBER--AFNO (ours)}         & 87.82 & \underline{86.26} & \underline{87.36} & 61.33 & 96.02 & 80.50 & 91.42 & 79.36 & 16.96 & 83.76 \\
\bottomrule
\end{tabular}}
\end{table*}

\begin{table*}[ht]
\centering
\caption{ Comparison on Synapse. AMBER-AFNO achieves the third best segmentation results (DSC), while being more efficient (Params in millions)}
\label{tab:syn_params}
\begin{tabular*}{\textwidth}{@{\extracolsep{\fill}}lcccc}
\toprule
\textbf{Methods} & \textbf{Params} & \textbf{ FLOPs} & \textbf{DSC} \\
\midrule
TransUNet\citep{TransUNet} & 96.07 & 88.91 & 77.49\\
UNETR\citep{UNETR} & 92.49 & 75.76 & 78.35\\
Swin\hbox{-}UNet\citep{Swin-Unet} & 62.83 & 384.2 & 83.48\\ 
nnFormer\citep{nnFormer}  &  150.5 &  213.4 & 86.57\\
UNETR++\citep{UNETR++}    & 42.96 & 47.98 &  \textbf{87.22} \\
LW-CTrans \citep{LW-CTrans}            & 4.42 & 275.92 & 73.34 \\
\midrule
\textbf{AMBER--AFNO (ours)} & {14.86 } & {161.24} & 83.76 \\
\bottomrule
\end{tabular*}
\end{table*}

\begin{figure*}
	\centering
	\includegraphics[width=.9\textwidth]{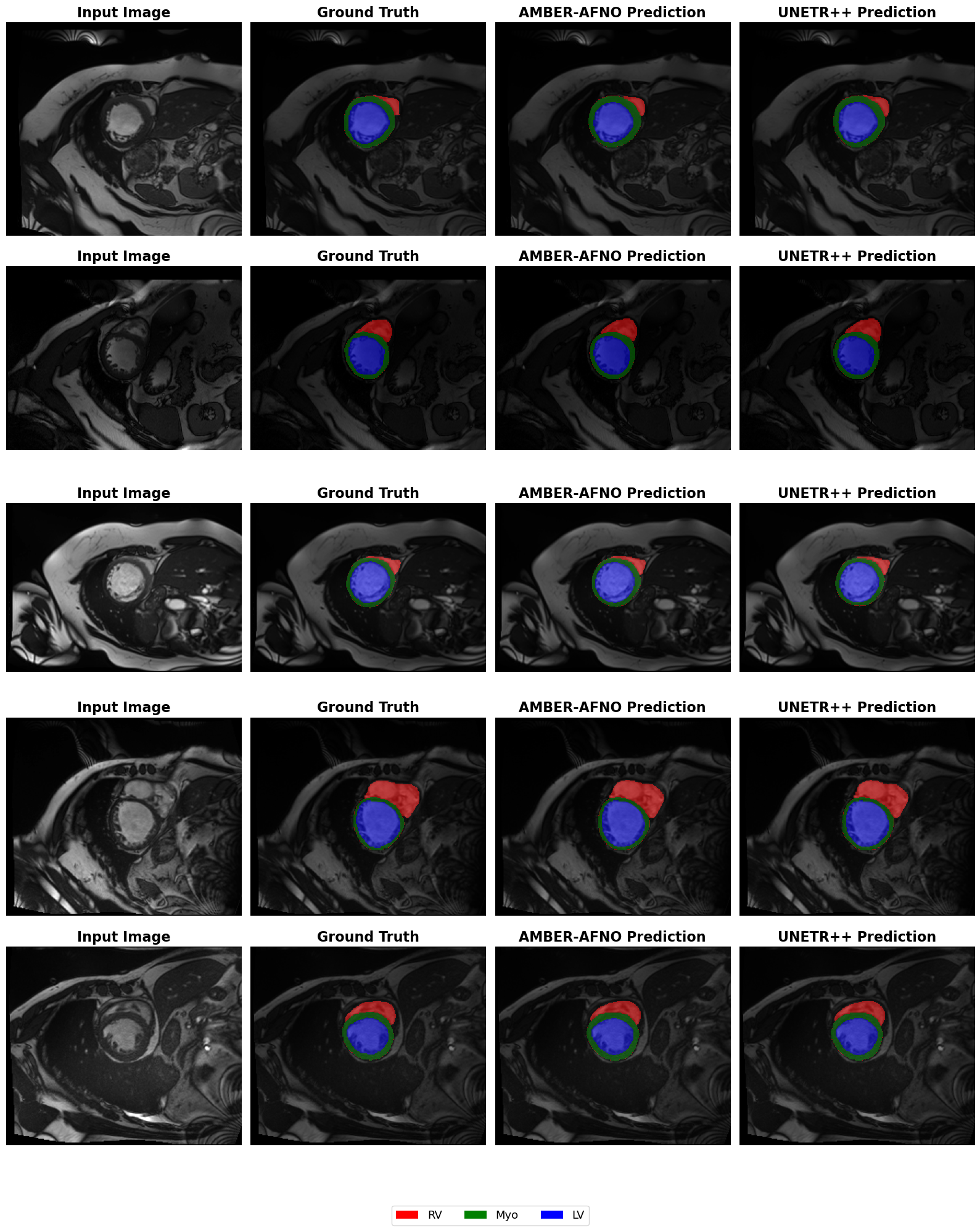}
	\caption{Qualitative comparison of AMBER-AFNO and UNETR++ segmentation predictions on representative ACDC validation samples. In the legend, \textbf{RV} denotes the Right Ventricular Cavity, \textbf{Myo} the Myocardium, and \textbf{LV} the Left Ventricular Cavity.}
	\label{acdc}
\end{figure*}
\subsection{Synapse Dataset}
\noindent According to Tab.~\ref{tab:synapse_results} and Tab.~\ref{tab:syn_params}, the proposed \textsc{AMBER-AFNO} yields an average Dice of $83.76\%$ on the Synapse validation set, which is ranked 3rd among the methods in comparison. Although UNETR++ and nnFormer gain higher Dice of $87.22\%$ and $86.57\%$, respectively, both require significantly more model parameters of $42.96$\,M and $150.5$\,M, respectively, in comparison to $14.86$\,M in \textsc{AMBER-AFNO}, indicating a good balance between accuracy and efficiency.\newline
Notably, compared to the state-of-the-art lightweight CNN–Transformer baseline LW-CTrans (73.34\%), AMBER-AFNO increases the Dice score by over 10 percentage points with comparable model compactness and computational complexity (161.24G vs. 275.92G FLOPs). This implies that frequency-domain global token mixing is more scalable than drastic convolution–attention compression when multiple anatomical classes and intricate organ boundaries are present.\newline
On  the Synapse dataset, the task is more difficult than ACDC and BraTS, as the model must learn the probability distributions of multiple abdominal organs with varying shapes and sizes. On this task, our AFNO-based encoder with fewer parameters still achieves consistent performance, demonstrating a favorable trade-off between representational capacity and computational cost for modeling global contextual information in the spectral domain. Therefore, this evidence further supports the use of AMBER-AFNO for 3D multi-organ segmentation under the low-latency constraint.

\begin{figure*}[p]
	\centering
	\includegraphics[width=.9\textwidth]{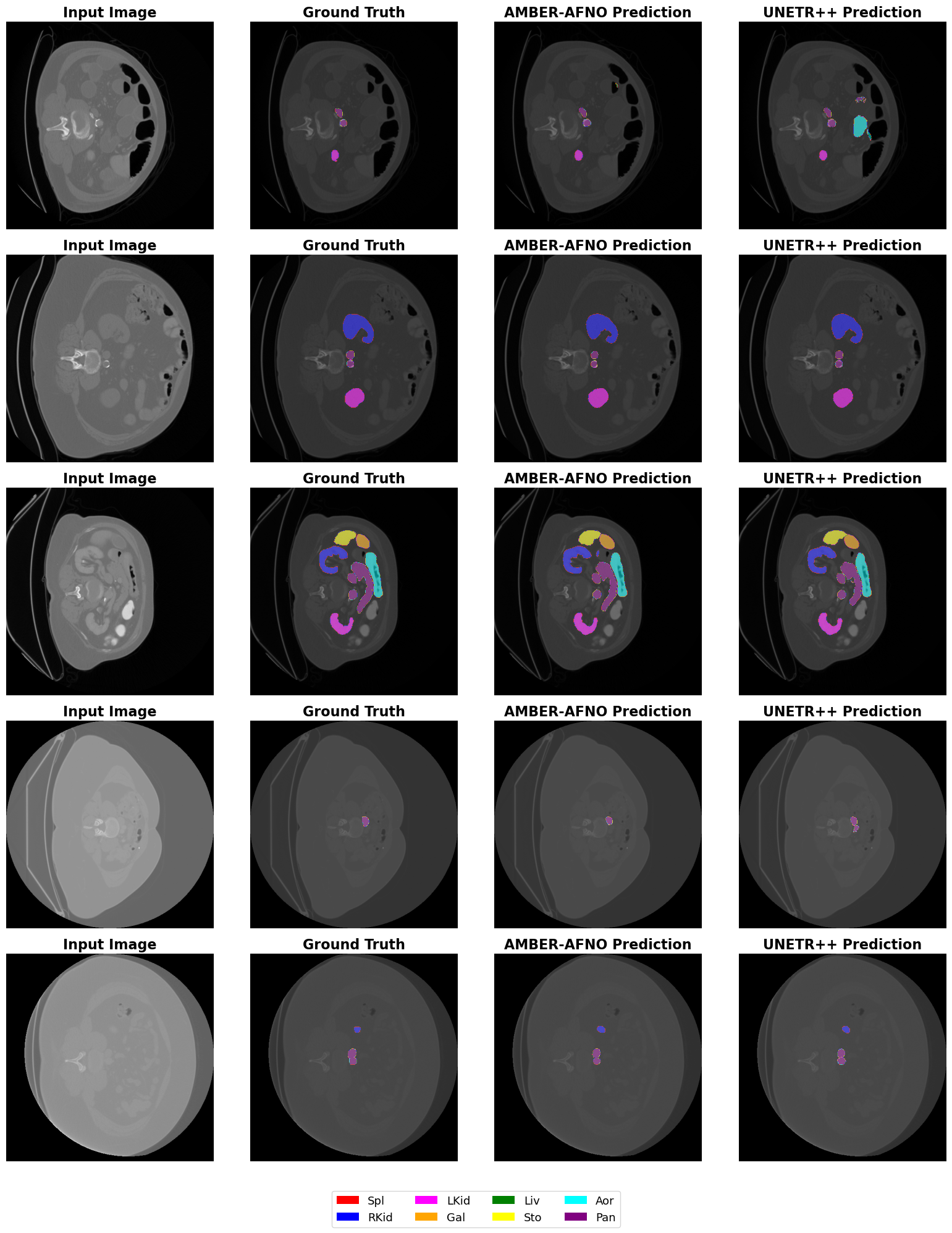}
	\caption{Visual Comparison of AMBER-AFNO and UNETR++ Model's Prediction on Synapse Dataset. In the legend, \textbf{Spl} denotes the Spleen, \textbf{RKid} the Right Kidney, \textbf{LKid} the Left Kidney, \textbf{Gal} the Gallbladder, \textbf{Liv} the Liver, \textbf{Sto} the Stomach, \textbf{Aor} the Aorta, and \textbf{Pan} the Pancreas. }
	\label{synapse}
\end{figure*}

\subsection{BraTS Dataset}
\noindent From Tab.~\ref{tab:brats_results}, we see that \textsc{AMBER-AFNO} yields the best mean Dice score on the BraTS validation set, with 82.82\%, which is slightly better than UNETR++ with 82.75\%. This is a strong result, as UNETR++ is one of the heavier transformer baselines that yield the best performance.\newline
Specifically, AMBER-AFNO achieves the highest Dice score (80.33\%) in the Enhancing Tumor (ET) region, which is generally the most challenging subregion due to its unclear boundary and limited spatial coverage. This implies that spectral-domain global mixing can model long-range contextual information while retaining the ability to detect detailed tumour structure.\newline
Compared with the lightweight CNN–Transformer baseline LW-CTrans (79.60\%), AMBER-AFNO achieved a more than 3\% increase in segmentation accuracy with a small number of parameters. This implies that frequency-domain token mixing works well for multimodal and multi-region segmentation tasks, further validating the generalizability of the proposed framework across various clinical applications.

\begin{table*}[ht]
\centering
\begingroup
\caption{Dice scores (\%) for Whole Tumor (WT), Enhancing Tumor (ET), Tumor Core (TC), and overall DSC (\%), together with HD95 (mm) on the BraTS validation set. \textbf{Bold} values denote the best result in each column, while \underline{underlined} values denote the second best.}
\label{tab:brats_results}
\begin{tabular*}{\textwidth}{@{\extracolsep{\fill}}lccccc}
\toprule
\textbf{Methods} 
& \textbf{WT} 
& \textbf{ET} 
& \textbf{TC} 
& \textbf{DSC} 
& \textbf{HD95} \\
\midrule

UNETR \citep{UNETR} 
& 90.35 & 76.30 & 77.02 & 81.22 & 6.61 \\

TransBTS \citep{TransBTS} 
& 90.91 & 77.86 & 76.10 & 81.62 & 5.80 \\

Swin-UNETR \citep{Swin-Unet} 
& 91.12 & 77.65 & \underline{78.41} & 82.39 & 5.33 \\

CoTr \citep{CoTr} 
& 91.01 & 77.52 & 77.43 & 81.99 & 5.78 \\

nnU-Net \citep{reference14} 
& 91.21 & 77.96 & 78.05 & 82.41 & 5.58 \\

nnFormer \citep{nnFormer} 
& \underline{91.23} & 77.84 & 77.91 & 82.34 & \underline{5.18} \\

UNETR++ \citep{UNETR++} 
& 91.27 & \underline{78.39} & \textbf{78.60} & \underline{82.75} & \textbf{5.05} \\

LW-CTrans \citep{LW-CTrans} 
& 89.58 & 73.83 & 75.38 & 79.60 & 6.91 \\

\midrule
\textbf{AMBER--AFNO (ours)} 
& 90.86 & \textbf{80.33} & 77.26 & \textbf{82.82} & 6.42 \\
\bottomrule
\end{tabular*}
\endgroup
\end{table*}

\begin{figure*}[ht]
	\centering
	\includegraphics[width=.9\textwidth]{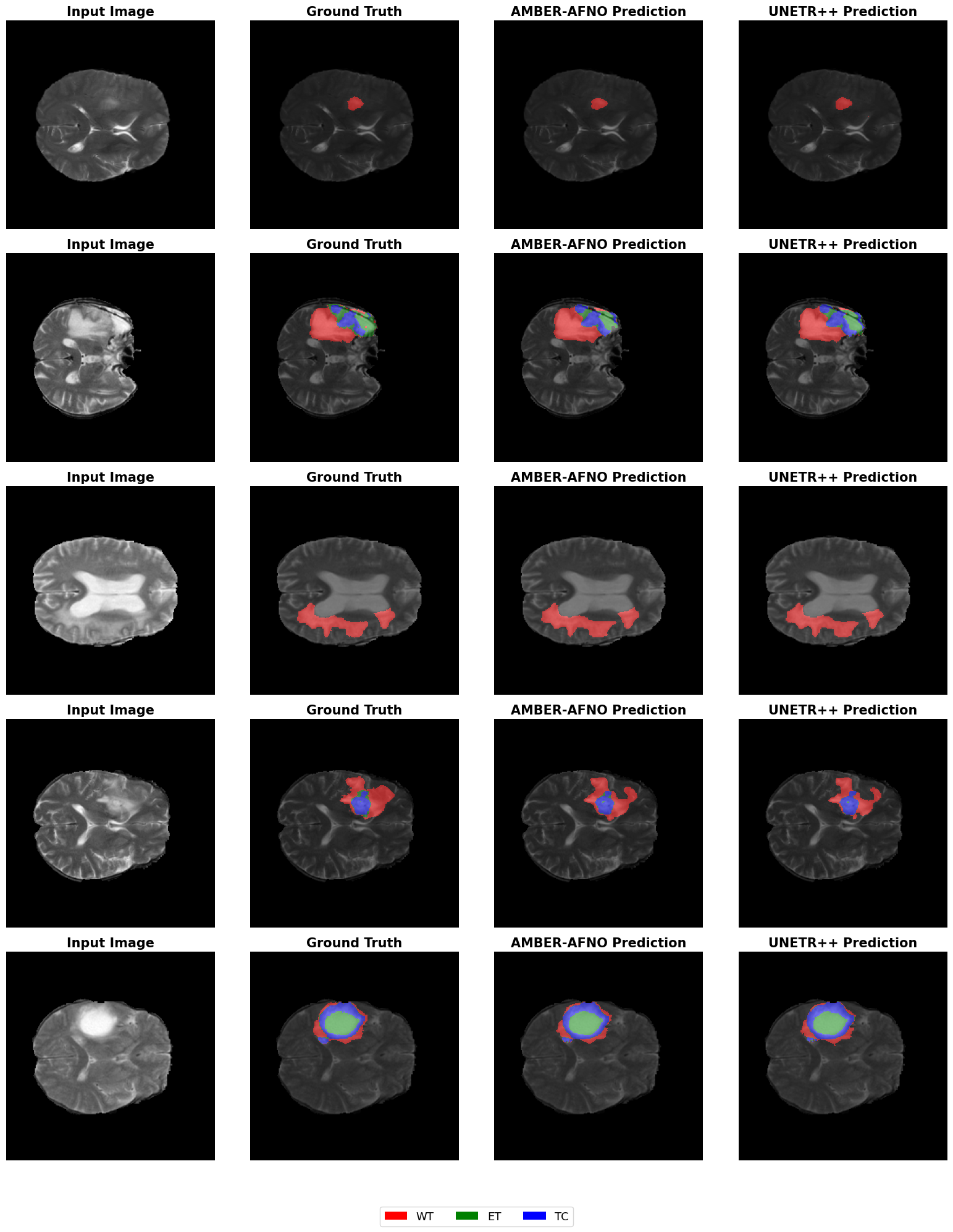}
	\caption{Visual comparison of AMBER-AFNO and UNETR++ predictions on the BraTS dataset. In the legend, \textbf{WT} denotes the Whole Tumor, \textbf{ET} the Enhancing Tumor, and \textbf{TC} the Tumor Core.}
	\label{brats}
\end{figure*}

\subsection{Discussion: Performance vs Efficiency Trade-off}

\noindent The experimental results in Tables ~\ref{tab:acdc_results}, ~\ref{tab:acdc_params}, ~\ref{tab:synapse_results}, ~\ref{tab:syn_params}, and ~\ref{tab:brats_results} clearly show that our AMBER-AFNO model consistently outperforms both heavy transformer models and recent lightweight CNN-Transformer models in key metrics.\vspace{2mm}

\noindent On the ACDC dataset, AMBER-AFNO achieves the best Dice Similarity Coefficient (92.85\%), which is superior to UNETR++ (92.83\%) by a small margin with much fewer parameters (14.77M vs. 81.55M). Compared to the light-weight model LW-CTrans (92.62\%), AMBER-AFNO improves the performance with much lower FLOPs (163.27G vs. 275.49G). This shows that the proposed spectral-domain token mixing could yield stronger representation power than current CNN-Transformer hybrids for lightweight models. \vspace{2mm} \newline 
On the multi-organ Synapse dataset, which consists of eight abdominal organs and higher anatomical diversity, AMBER-AFNO obtains a mean Dice of 83.76, ranking in third place. Though larger models, such as UNETR++ and nnFormer, achieve higher absolute Dice scores, they require significantly more model parameters. Furthermore, relative to LW-CTrans (73.34), AMBER-AFNO significantly improves performance by more than 10 percentage points with a lower parameter count. Meanwhile, it incurs lower computational cost (161.24G vs. 275.92G FLOPs), indicating that frequency-domain token mixing is more suitable than pure convolution-attention fusion for multi-class segmentation.\vspace{2mm} \newline 
On the challenging BraTS dataset with high intra-class variability and multi-region tumor segmentation task, the proposed AMBER-AFNO gains the best average Dice (82.82\%) among all methods. Specifically, it outperforms the recent UNETR++ (82.75\%) while using a much smaller model. Meanwhile, compared with the lightweight counterpart LW-CTrans (79.60\%), the proposed model improves Dice accuracy while remaining lightweight. Moreover, our method achieves the highest Dice score (80.33\%) on the Enhancing Tumor dataset, further verifying its superiority in identifying fine-grained tumor substructures.

\noindent From these three comparisons, the one against LW-CTrans is especially interesting because, although optimized for the fewest parameters and lowest computational cost, LW-CTrans always yields lower segmentation Dice scores than AMBER-AFNO. Instead, AMBER-AFNO maintains a moderate model size while always offering higher Dice scores on ACDC, Synapse, and BraTS.

\noindent This performance-efficiency trade-off demonstrates the effectiveness of our central hypothesis. Specifically, AMBER-AFNO removes token-to-token interactions and retains only global context modeling by using Adaptive Fourier Neural Operators instead of the quadratic self-attention mechanism, leading to a remarkable decrease in parameter growth and computational cost, without compromising segmentation accuracy.

\noindent All this is obtained without resorting to decoder branches, multi-scale fusion hierarchies, or deep stacks of attention modules. The overall architecture's simplicity and modularity allow straightforward deployment across different datasets with negligible task-specific adaptations. All in all, these characteristics make AMBER-AFNO a workable and scalable solution compared to heavy transformer-based models and existing light CNN-Transformer models.

\noindent 
Tab.~\ref{tab:amber_comput} reports the memory profile and the latency of 
AMBER-AFNO across four hardware configurations combining two modern GPUs 
(NVIDIA H100 and NVIDIA L40) and two different CPUs (AMD EPYC 7413-24c and AMD EPYC 9534-64c). 
AMBER-AFNO exhibits an extremely light memory footprint, 
requiring only 2.96\,GB of GPU memory for full-resolution 3D inference, making 
the model deployable even on mid-range or shared accelerator environments. 
In terms of speed, the model achieves sub-100\,ms latency on the NVIDIA L40 and 
remains below 160\,ms on the H100, demonstrating high throughput for the BraTS
$128^3$ volumetric inputs. CPU-only execution is naturally slower, but still 
processes each volume within 3–3.5\,s, which remains practical for 
non-real-time clinical workflows. Overall, the measurements highlight the 
efficiency of Amber-AFNO, combining low memory demand with fast inference 
suitable for scalable deployment.

\begin{table*}[ht]

\centering
\caption{Computational performance of Amber-AFNO on the BraTS tumor input 
($1\times128\times128\times128$). GPU T.\ and CPU T.\ denote the average forward-pass 
latency over 20 runs. Mem indicates the peak GPU memory usage during inference.}
\label{tab:amber_comput}
\begin{tabular}{lccccccc}
\toprule
GPU & CPU & Params (M) & FLOPs (G) & Mem (GB) & GPU T. (ms) & CPU T. (ms) \\
\midrule
NVIDIA H100 PCIe & AMD EPYC 7413 (24c) & 9.84 & 122.91 & 2.96 & 159.6 & 3305.0 \\
NVIDIA L40       & AMD EPYC 9534 (64c) & 9.84 & 122.91 & 2.96 &  84.2 & 3453.2 \\
\bottomrule
\end{tabular}
\end{table*}

\noindent As discussed in Section \ref{sec:ablation_acdc}, along with the AMBER-ANFO, we have also performed an ablation study on the lightest version of AMBER-AFNO with embedding dimension [32, 64, 128, 256].  Fig. \ref{performance_plot} shows the efficiency–accuracy trade-off of different models on the ACDC dataset. The mean Dice score (DSC) is plotted against the total number of parameters. AMBER-AFNO achieves the highest DSC (92.85\%) while maintaining a moderate number of parameters. AMBER-AFNO (light) and UNETR++ both achieve 92.83\%, but AMBER-AFNO (light) uses significantly fewer parameters. Although LW-CTrans has fewer parameters, it does not surpass AMBER-AFNO or AMBER-AFNO (light) in performance.

\section{Ablation Study}
\label{sec:ablation_acdc}
\noindent
In order to better demonstrate the efficiency of the proposed AFNO, we make the encoder variants trained on ACDC dataset share all hyper-parameters including network depth, number of heads, MLP expansion ratio, deep supervision (see details in Section~\S\ref{Hierarchical Transformer Encoder}), as well as embedding dimension \([32,\,64,\,128,\,256]\). The only difference between these encoders is the token-mixing block: for \textbf{AMBER (MHSA)}, the AFNO blocks are replaced with the conventional multi-head self-attention (MHSA); for \textbf{AMBER-AFNO (light)}, we use the proposed AFNO block with the above-mentioned setting.

\begin{figure}
	\centering
	\includegraphics[width=\columnwidth]{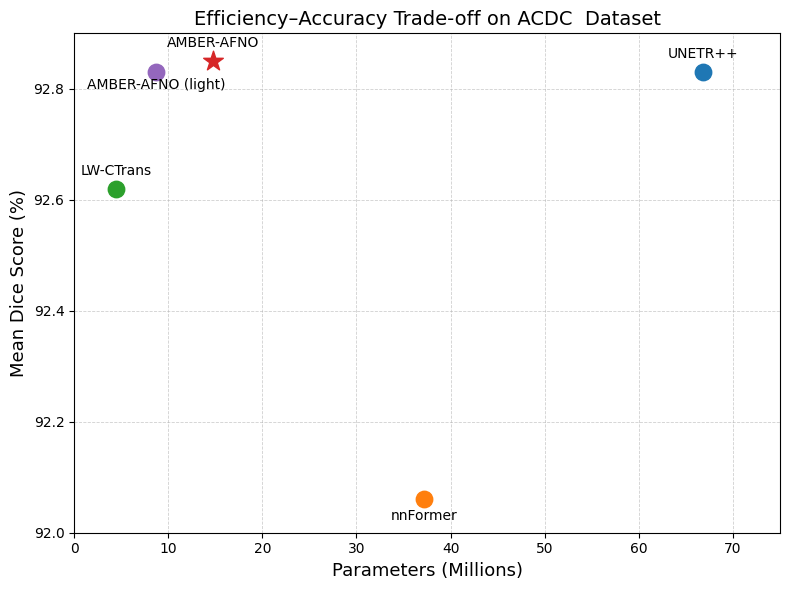}
	\caption{Efficiency–accuracy trade-off of different models on the ACDC  Dataset}
        \label{performance_plot}
\end{figure}

\smallskip
\noindent Despite having less than half of the parameters and the FLOPs count of the MHSA module counterpart (\textbf{8.7M, 58.29 GFLOPs}) versus (\textbf{19.01M, 132.07 GFLOPs}), \emph{AMBER-AFNO (light)} achieves a Dice score of \textbf{92.83\%}, which is on par with the significantly larger \textsc{UNETR++} (66.8M parameters). \newline
However, the MHSA-based model achieves a relatively lower mean Dice score of \textbf{92.03\,\%}, representing a performance decrease of about 0.8\,\% while being more complex. Therefore, the superiority of spectral token mixing over the quadratic self-attention is straightforwardly illustrated here.
Notably, a comparison with the latest, lightweight CNN-Transformer model, \textbf{LW-CTrans}, provides a clearer view of the trade-off between accuracy and computational complexity. Though LW-CTrans performs well in terms of the number of parameters, \textbf{4.42M}, it suffers from significantly higher computational complexity, \textbf{275.49 GFLOPs}, and a decreased Dice score of 92.62. Specifically, the proposed \textbf{AMBER-AFNO (light)} enjoys a higher Dice score of 92.83 with almost 5$\times$ lower FLOPs compared to LW-CTrans, indicating a better performance-complexity ratio. AMBER-AFNO achieves the highest Dice score of 92.85\% while maintaining a substantially smaller parameter count than UNETR++, although with higher FLOPs due to the full 3D spectral mixing.\newline
The number of parameters, FLOPs, and DSC of all models are listed in Tab. \ref{tab:ablation}. In general, these results demonstrate that the proposed AFNO-based frequency-domain token mixing offers a better trade-off between efficiency and accuracy than conventional MHSA and recent CNN/transformer models for 3-D medical image segmentation. Fig. \ref{performance_plot} visually demonstrates how AMBER-AFNO performs better compared to other state-of-the-art models in terms of dice score and total number of parameters. 

\smallskip
\noindent In order to test the robustness and reproducibility of the proposed model based on the AFNO, we conducted a sensitivity analysis where we tested different scenarios, including the impact of the number of epochs, learning rate, batch size, embedding dimension, dropout ratio, number of blocks in each stage, and deep supervision. The outcomes of the sensitivity analysis are summarized in the table. We can see from the table that the Dice score remains within the range of 0.91-0.93 for a wide range of hyperparameters, which suggests that the proposed model is not sensitive to the hyperparameters (within a reasonable range). Using more blocks or a relatively larger embedding dimension can yield marginal improvements in the Dice score, but it will also increase computational cost. In addition, deep supervision or a relatively small learning rate can speed up the training but has little impact on the Dice score. These results suggest that the proposed model is robust and can be easily generalized to other datasets (i.e., we do not need to re-tune many hyperparameters).

\smallskip
\noindent \textit{Notes on parameters:}
\textbf{Deep Supervision} indicates whether additional prediction heads are used, enabling loss computation at multiple resolutions.
\textbf{Batch Size} denotes the number of 3D volumes processed per iteration;
\textbf{Epochs} represents total training cycles;
\textbf{LR} is the initial learning rate;
\textbf{Embed Dim.} lists the embedding dimensions across encoder stages;
\textbf{Dropout} specifies the dropout probability applied in the MLP layers;
and \textbf{Blocks/Stage} refers to the number of AFNO blocks used in each hierarchical encoder stage.

\begin{table*}[ht]
    \centering
    \caption{Efficiency–accuracy trade-off on the ACDC validation set. AFNO delivers state-of-the-art Dice scores with markedly fewer parameters. Best results are highlighted in \textbf{bold}, and second-best results are \underline{underlined}.}
    \label{tab:ablation}
    \begin{tabular*}{\textwidth}{@{\extracolsep{\fill}}lccc}
        \toprule
        \textbf{Method}              & \textbf{Params} & \textbf{FLOPs} & \textbf{DSC (\%)} \\ 
        \midrule
        UNETR++\citep{UNETR++}       & 66.8 & 43.71 & \underline{92.83} \\ 
        nnFormer\citep{nnFormer}        & 37.20 & 399.25 & {92.06} \\ 
        LW-CTrans \citep{LW-CTrans}           & 4.42 & 275.49 & 92.62 \\
        AMBER-AFNO                  & {14.77} & 163.27 & \textbf{92.85} \\ 
        AMBER-AFNO (light)          & 8.70 & 58.29  & \underline{92.83} \\ 
        AMBER (MHSA)                & 19.01 & 132.07 & 92.03 \\ 
        \bottomrule
    \end{tabular*}
\end{table*}

\begin{table*}[ht]
    \centering
    \caption{Sensitivity analysis of key hyperparameters on the ACDC validation set. The model demonstrates stable Dice scores across a wide range of configurations, confirming robustness to variations in training strategy and architecture depth.}
    \label{tab:sensitivity}
    \begin{tabular*}{\textwidth}{@{\extracolsep{\fill}}lcccccccc}
        \toprule
        \textbf{Experiments} & \textbf{Deep Supervision} & \textbf{Batch Size} & \textbf{Epochs} & \textbf{LR} & \textbf{Embed Dim.} & \textbf{Dropout} & \textbf{Blocks/Stage} & \textbf{Dice} \\ 
        \midrule
        1 & True  & 4 & 1000 & 0.0100 & [32, 64, 128, 256] & 0.1 & [2, 2, 4, 2] & \textbf{0.93}\\
        2 & False & 1 & 1500 & 0.0037 & [32, 64, 128, 256] & 0.1 & [3, 4, 6, 3] & 0.92 \\
        3 & True  & 4 & 1000 & 0.0100 & [64, 128, 320, 512] & 0.0 & [3, 4, 6, 3] & \textbf{0.93} \\
        4 & False & 4 & 1000 & 0.0100 & [64, 128, 320, 512] & 0.0 & [3, 4, 6, 3] & 0.91 \\
        5 & False & 4 & 1000 & 0.0100 & [64, 128, 320, 512] & 0.0 & [3, 4, 6, 3] & \textbf{0.93}\\
        6 & False & 4 & 1000 & 0.0053 & [64, 128, 320, 512] & 0.0 & [3, 6, 36, 3] & 0.92 \\
        \bottomrule
    \end{tabular*}
\end{table*}

\section{Conclusions}\label{Conclusion}
\noindent In this study, we introduced AMBER-AFNO, a novel framework for 3D medical image segmentation. It uses Adaptive Fourier Neural Operators (AFNO) instead of quadratic self-attention. Unlike other approaches that approximate or sparsify attention, we directly reformulate the problem of modeling global context. We do this by removing token-to-token interactions and ensuring tokens are mixed in the spectral domain. Our design delivers quasi-linear complexity and linear memory growth with respect to volume size. This effectively overcomes a major drawback of transformer-based 3D segmentation.

\noindent To the best of our knowledge, no prior work has integrated AFNO-based spectral token mixing into a transformer framework for 3D medical image segmentation. Other efficient CNN–Transformer variants mainly seek to reduce the complexity of attention or convolutional blocks. In contrast, AMBER-AFNO removes quadratic self-attention from the transformer encoder. It still preserves global receptive-field modeling through frequency truncation.

\noindent On three public benchmarks (ACDC, Synapse, BraTS), AMBER-AFNO matches or surpasses heavy transformer models (UNETR++ \citep{UNETR++}, nnFormer \citep{nnFormer}) with substantially fewer parameters (nearly 78\% less on the ACDC dataset). Compared to the light CNN-Transformer LW-CTrans \citep{LW-CTrans}, AMBER-AFNO achieves higher Dice scores. It also has a smaller model size and lower FLOPs on ACDC and Synapse. This demonstrates a more favorable balance between accuracy and efficiency versus quadratic or compressed attention-based methods.

\noindent Although performance on the Synapse dataset suggests that further enhancement of local structural modeling may be beneficial, the proposed framework establishes a new direction for efficient volumetric segmentation: global context modeling through spectral operators rather than attention matrices. Future work will explore hybrid spectral–spatial strategies, improved multi-scale integration, and transfer learning for domain adaptation.

\noindent In summary, AMBER-AFNO bridges the gap between computationally intensive transformer architectures and lightweight CNN–Transformer models by introducing an alternative global information-mixing strategy. Its compact parameterization and scalable complexity make it well-suited for deployment in resource-constrained clinical environments.

\section*{Acknowledgments}
\noindent This work has been funded by project code \texttt{PIRO1\_00011} ``IBISCo'', PON 2014--2020, for all three entities (INFN, UNINA, and CNR). \newline
\newline
We acknowledge the use of the ADHOC (Astrophysical Data HPC Operating Center) resources, within the
project "Strengthening the Italian Leadership in ELT and SKA (STILES)", proposal nr. IR0000034, admitted and
eligible for funding from the funds referred to in the D.D. prot. no. 245 of August 10, 2022 and D.D. 326 of
August 30, 2022, funded under the program "Next Generation EU" of the European Union, “Piano Nazionale
di Ripresa e Resilienza” (PNRR) of the Italian Ministry of University and Research (MUR), “Fund for the creation
of an integrated system of research and innovation infrastructures”, Action 3.1.1 "Creation of new IR or
strengthening of existing IR involved in the Horizon Europe Scientific Excellence objectives and the
establishment of networks".


\bibliographystyle{cas-model2-names}

\bibliography{cas-refs_no_marks}

\end{document}